\documentclass[12pt,a4paper,reqno]{article}
\usepackage{graphics}
\usepackage{blkarray}

\usepackage{color}

\usepackage{epsfig}

\usepackage{amsmath,amsthm,amssymb}
\newtheorem{theorem}{Theorem}[section]

\newtheorem{lemma}[theorem]{Lemma}

\newtheorem{example}[theorem]{Example}
\theoremstyle{definition}

\theoremstyle{remark}

\numberwithin{equation}{section}

\begin{document}
\title{  On Zero-Divisor Graph of the ring $\mathbb{F}_p+u\mathbb{F}_p+u^2 \mathbb{F}_p$}

\author{N. Annamalai\\
Faculty On Contract\\
Department of Mathematics\\
National Institute of Technology Puducherry\\ Karaikal,
 India\\
{Email: algebra.annamalai@gmail.com}
}
\date{}
\maketitle
\begin{abstract}
In this article, we discussed the zero-divisor graph of a commutative ring with identity $\mathbb{F}_p+u\mathbb{F}_p+u^2 \mathbb{F}_p$ where $u^3=0$ and $p$ is an odd prime.
We find the clique number, chromatic number, vertex connectivity, edge connectivity, diameter and girth of a zero-divisor graph associated with the ring. We find some of topological indices and the main parameters of the code derived from the incidence matrix of the zero-divisor graph $\Gamma(R).$ Also, we find the eigenvalues, energy and spectral radius  of both adjacency and Laplacian matrices of $\Gamma(R).$
\end{abstract}



The zero-divisor graph has attracted a lot of attention in the last few years. In 1988, Beck \cite{beck} introduced the zero-divisor graph.  He included the additive identity of a ring $R$ in the definition and was mainly interested in the coloring of commutative rings.
Let $\Gamma$ be a simple graph whose vertices are the set of zero-divisors of the ring $R,$ and two distinct vertices are adjacent if the product is zero. Later it was modified by Anderson and Livingston \cite{and}. They redefined the definition  as a simple graph that only considers the non-zero zero-divisors of a commutative ring $R.$ 

Let $R$ be a commutative ring with identity and $Z(R)$ be the set of zero-divisors of $R.$ The zero-divisor graph $\Gamma(R)$ of a ring $R$ is an undirected graph whose vertices are the non-zero zero-divisors of $R$ with two distinct vertices $x$ and $y$ are adjacent if and only if $xy=0.$ 
In this article, we consider the zero-divisor graph $\Gamma(R)$ as a graph with vertex set $Z^*(R)$  the set of non-zero zero-divisors of the ring $R.$ 
Many researchers are doing research in this area \cite{amir, kavaskar, red}.  

Let $\Gamma=(V, E)$ be a simple undirected graph with vertex set $V,$ edge set $E.$ An incidence matrix of a graph
$\Gamma$ is a $|V|\times|E|$ matrix $Q(\Gamma)$ whose rows are labelled by the vertices and 
columns by the edges and entries $q_{ij}=1$ if the vertex labelled by row $i$ 
is incident with the edge labelled by column $j$ and $q_{ij}=0$ otherwise.

The adjacency matrix $A(\Gamma)$ of the graph $\Gamma,$ is the $|V|\times |V|$ matrix defined as follows. The rows and
the columns of $A(\Gamma)$ are indexed by $V.$ If $i \neq j$ then the $(i, j)$-entry of $A(\Gamma)$ is
0 for vertices $i$ and $j$ nonadjacent, and the $(i, j)$-entry is 1 for $i$ and $j$ adjacent. The
$(i,i)$-entry of $A(\Gamma)$ is 0 for $i = 1, \dots, |V|.$ 
For any (not necessarily bipartite) graph $\Gamma,$ the energy of the graph is defined as
$$\varepsilon(\Gamma) =\sum\limits_{i=1}^{|V|} |\lambda_i|,$$ where $\lambda_1, \dots,\lambda_{|V|}$ are the eigenvalues of  $A(\Gamma)$ of $\Gamma.$

The Laplacian matrix $L(\Gamma)$ of $\Gamma$  is the $|V|\times |V|$ matrix defined as follows. The rows and
columns of $L(\Gamma)$ are indexed by $V.$ If $i\neq j$ then the $(i, j)$-entry of $L(\Gamma)$ is 0 if vertex $i$ and $j$ are not adjacent, and it is $-1$ if $i$ and $j$ are adjacent. The $(i,i)$-entry of $L(\Gamma)$ is $d_i$, the degree of the vertex $i,$ $i = 1, 2, \dots, |V|.$
Let $D(\Gamma)$ be the diagonal matrix of vertex degrees. If $A(\Gamma)$ is the adjacency matrix of $\Gamma,$ then note that $L(\Gamma) = D(\Gamma)-A(\Gamma).$
Let $\mu_1, \mu_2,\dots,\mu_{|V|}$ are eigenvalues of $L(\Gamma).$ Then
the Laplacian energy $LE(\Gamma)$ is given by $$LE(\Gamma)=\sum\limits_{i=1}^{|V|} \Big|\mu_i-\frac{2|E|}{|V|}\Big|.$$

\begin{lemma}\cite{bapat}\label{a}
Let $\Gamma = (V, E)$ be a graph, and let $0 = \lambda_1 \leq \lambda_2 \leq \dots \leq \lambda_{|V|}$ be the eigenvalues of
its Laplacian matrix $L(\Gamma).$ Then, $\lambda_2 > 0$ if and only if $\Gamma$ is connected.
\end{lemma}

The Wiener index of a connected graph $\Gamma$ is defined as the sum of distances between each pair of vertices, i.e.,
$$W(\Gamma)=\sum_{\substack{a, b \in V\\ a \neq b}}d(a, b),$$
where $d(a, b)$ is the length of shortest path joining $a$ and $b.$ 

The degree of $v\in V,$ denoted by $d_v,$ is the number of vertices adjacent to $v.$

The Randi\'{c} index (also known under the name connectivity index) is a much investigated degree-based topological index. It was invented in 1976 by Milan Randi\'{c} \cite{randic} and is defined as $$R(\Gamma)=\sum_{(a,b)\in E} \frac{1}{\sqrt{d_a d_b}}$$
with summation going over all pairs of adjacent vertices of the graph.

The Zagreb indices were introduced more than thirty years ago by Gutman and Trinajesti\'{c} \cite{gutman}.
 For a graph $\Gamma$, the first Zagreb index $M_1(\Gamma)$  and the second Zagreb index $M_2(\Gamma)$ are, respectively, defined as follows:
 $$M_1(\Gamma)=\sum_{a\in V} d_a^2$$
 $$M_2(\Gamma)=\sum_{(a,b)\in E}d_a d_b.$$
An edge-cut of a connected graph $\Gamma$ is the set $S\subseteq E$ such that $\Gamma- S=(V, E-S)$ is disconnected.
The edge-connectivity $\lambda(\Gamma)$ is the minimum cardinality of an edge-cut. 
The minimum $k$ for which there exists a $k$-vertex cut is called the vertex connectivity or simply the connectivity of $\Gamma$ it is denoted by $\kappa(\Gamma).$

For any connected graph $\Gamma,$ we have $\lambda(\Gamma)\leq \delta(\Gamma)$
where $\delta(\Gamma)$ is minimum degree of the graph $\Gamma.$

The chromatic number of a graph $\Gamma$ is the minimum number of colors needed to color the vertices of $\Gamma$ so that adjacent vertices of $\Gamma$ receive distinct colors and is denoted
by $\chi(\Gamma).$ 
The clique number of a graph $\Gamma$ is the maximum size of a subset $C$ of $V$ for which $xy = 0,$ for all $x, y \in C$ and it is denoted by $\omega(\Gamma).$ That means, $\omega(\Gamma)$ is the maximum size of a complete subgraph of $\Gamma.$ Note that for any graph $\Gamma,$ $\omega(\Gamma) \leq \chi(\Gamma).$ 

 Beck\cite{beck} conjectured that if $R$ is a finite chromatic ring, then $\omega(\Gamma(R))=\chi(\Gamma(R))$ where $\omega(\Gamma(R)), \chi(\Gamma(R))$ are the clique number and the chromatic number of $\Gamma(R)$, respectively. He also verified that the conjecture is true for several examples of rings. Anderson and Naseer, in \cite{and},
disproved the above conjecture with a counterexample. $\omega(\Gamma(R))$ and $\chi(\Gamma(R))$ of the zero-divisor graph associated to the ring $\mathbb{F}_p+u\mathbb{F}_p+u^2 \mathbb{F}_p$ are same. For basic graph theory, one can refer \cite{R.B, bapat}.

 Let $\mathbb{F}_q$ be a finite field with $q$ elements. 
Let $x=(x_1, \dots, x_n)\in \mathbb{F}_q^n,$ then the Hamming weight $w_{H}(x)$ of $x$ is defined by the number of non-zero coordinates in $x.$ 
Let 
$x=(x_1,\dots, x_n), y = (y_1, \dots, y_n) \in \mathbb{F}_q^n,$ the 
Hamming distance  $d_H(x,y)$ between $x$ and $y$ is defined by the number of coordinates in which they differ.

A $q$-ary code of length $n$ is a non-empty subset $C$ of $\mathbb{F}_{q}^{n}.$  If $C$ is a subspace of $\mathbb{F}_{q}^{n},$ then $C$ is called a $q$-ary linear code of length $n.$ An element of  $C$ is called a \emph{codeword}. The minimum Hamming distance of a code $C$ is defined by
$$
d_{H}(C)=\min\{
d_{H}(c_{1}, 
c_{2}) \mid c_1\neq c_2, {c_{1},c_{2}\in C}\}.$$
The minimum weight $w_{H}(C)$ of a code $C$ is the smallest among all weights of the non-zero codewords of $C.$ For $q$-ary linear code, we have 
$d_{H}(C)=w_{H}(C).$ For basic coding theory, we refer \cite{san}.

A linear code of length $n,$ dimension $k$ and minimum
distance $d$  is denoted by $[n, k, d]_{q}.$  
The code generated by the rows of the incidence matrix $Q(\Gamma)$ of the  graph $\Gamma$ is denoted by $C_p(\Gamma)$ over the finite field $\mathbb{F}_p.$
\begin{theorem}\cite{dan}\label{21}
	\begin{itemize}
		\item[1.] Let $\Gamma = (V, E)$ be a connected graph and let  $G$ be a $|V|\times|E|$ incidence matrix for $\Gamma.$ Then, the main parameters of the code	$C_2(G)$ is $[|E|, |V|- 1, \lambda(\Gamma)]_2.$
		\item[2.]  Let $\Gamma = (V, E)$ be a connected bipartite graph and let $G$ be a $|V|\times|E|$ incidence matrix for $\Gamma.$ Then the incidence matrix generates 
		$[|E|, |V|-1,\lambda(\Gamma)]_p$ code for odd prime $p.$
	\end{itemize}
\end{theorem}
Codes from the row span of incidence matrix or adjacency matrix of various  graphs are studied in \cite{anna,malai,dan, cd1, cd2}.

Let $p$ be an odd prime.  The ring $\mathbb{F}_p+u\mathbb{F}_p+u^2 \mathbb{F}_p$ is defined as a characteristic $p$ ring subject to restrictions $u^3=0.$ The ring isomorphism
$\mathbb{F}_p+u\mathbb{F}_p+u^2 \mathbb{F}_p \cong \frac{\mathbb{F}_p[x]}{\langle x^3\rangle}$
is obvious to see. An element $a+ub+u^2 c\in R$ is unit if and only if $a\neq 0.$

Throughout this article, we denote the ring $\mathbb{F}_p+u\mathbb{F}_p+u^2 \mathbb{F}_p$ by $R.$ In this article, we discussed the zero-divisor graph of a commutative ring with identity $\mathbb{F}_p+u\mathbb{F}_p+u^2 \mathbb{F}_p$ where $u^3=0$
and we find the clique number, chromatic number,  vertex connectivity, edge connectivity, diameter, and girth of the graph $\Gamma(R),$ in Section 2. In Section 3, we find some of topological indices of $\Gamma(R).$  In Section 4,  we find the main parameters of the code derived from incidence matrix of the zero-divisor graph $\Gamma(R).$  Finally, We find the eigenvalues, energy and spectral radius of both adjacency and Laplacian matrices in Section 5.

\section{Zero-divisor graph $\Gamma(R)$ of the ring $R$}
In this section, we discuss the zero-divisor graph $\Gamma(R)$ of the ring $R$ and we find the clique number, chromatic number, vertex connectivity, edge connectivity, diameter, and girth of the graph $\Gamma(R).$

Let
$A_u=\{x u\mid x\in \mathbb{F}_p^{*}\},$
$A_{u^2}=\{x u^2\mid x\in \mathbb{F}_p^{*}\}$ and
$A_{u+u^2}=\{x u+y u^2\mid x, y\in \mathbb{F}_p^{*}\}.$ Then $|A_u|=(p-1),$ $|A_{u^2}|=(p-1)$ and $|A_{u+u^2}|=(p-1)^2.$
Therefore, $Z^{*}(R)=A_u\cup A_{u^2}\cup A_{u+u^2}$ and $|Z^{*}(R)|=|A_u|+|A_{u^2}|+|A_{u+u^2}|=(p-1)+(p-1)+(p-1)^2=p^2-1.$
\begin{figure}
  \begin{center}
\includegraphics{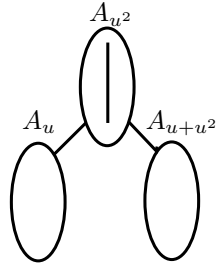}
  \end{center}
  \caption{Zero-divisor graph of $R=\mathbb{F}_p+u\mathbb{F}_p+u^2\mathbb{F}_p$}
\end{figure}
As $u^3=0,$ every vertices of $A_u$ is adjacent with every vertices of $A_{u^2},$ every vertices of $A_{u^2}$ is adjacent with every vertices of $A_{u+u^2}$ and any two distinct vertices of $A_{u^2}$ are adjacent. From the diagram, the graph $\Gamma(R)$ is connected with $p^2-1$ vertices and $(p-1)^2+(p-1)^3+\frac{(p-1)(p-2)}{2}=\frac{1}{2}(2p^3-3p^2-p+2)$ edges.

\begin{example}\label{a}
For $p=3,$ $R=\mathbb{F}_3+u\mathbb{F}_3+u^2\mathbb{F}_3.$ Then
$A_u=\{u, 2u\},$ $A_{u^2}=\{ u^2, 2u^2\},$ $A_{u+u^2}=\{ u+u^2, 2u+2u^2, u+2u^2, 2u+u^2\}.$
\begin{figure}
  \begin{center}
\includegraphics{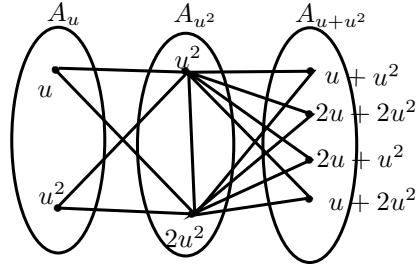}
  \end{center}
  \caption{Zero-divisor graph of $R=\mathbb{F}_3+u\mathbb{F}_3+u^2\mathbb{F}_3$}
\end{figure}
The number of vertices is 8 and the number of edges is 13.
\end{example}

\begin{theorem}
The diameter of the zero-divisor graph $diam(\Gamma(R))=2.$
\end{theorem}
\begin{proof}
From the Figure 1, we can see that the distance between any two distinct vertices are either 1 or 2. Therefore, the maximum of distance between any two distinct vertices is 2. Hence, $diam(\Gamma(R))=2.$
\end{proof}
\begin{theorem}
The clique number $\omega(\Gamma(R))$ of $\Gamma(R)$ is $p.$
\end{theorem}
\begin{proof}
From the Figure 1, $A_{u^2}$ is a complete subgraph(clique) in $\Gamma(R).$ If we add exactly one vertex $v$ from either $A_u$ or $A_{u+u^2},$ then resulting subgraph form a complete subgraph(clique). Then $A_{u^2}\cup\{v\}$ forms a complete subgraph with maximum vertices. Therefore, the clique number of $\Gamma(R)$ is $\omega(\Gamma(R))=|A_{u^2}\cup\{v\}|=p-1+1=p.$
\end{proof}

\begin{theorem}
The chromatic number $\chi(\Gamma(R))$ of $\Gamma(R)$ is $p.$
\end{theorem}
\begin{proof}
Since $A_{u^2}$ is a complete subgraph with $p-1$ vertices in $\Gamma(R),$ then at least $p-1$ different colors needed to color the vertices of $A_{u^2}.$ And no two vertices in $A_{u}$ are adjacent then one color different from previous $p-1$ colors is enough to color all vertices in $A_u.$ We take the same color in $A_u$ to color vertices of $A_{u+u^2}$ as there is no direct edge between $A_u$ and $A_{u+u^2}.$   
Therefore, minimum $p$ different colors required for proper coloring. Hence, the chromatic number $\chi(\Gamma(R))$ is $p.$
\end{proof}

The above two theorems show that the clique number and the chromatic number of our graph are same. 

\begin{theorem}
The girth of the graph $\Gamma(R)$ is 3.
\end{theorem}
\begin{proof}
We know that the girth of a complete graph is 3. From the Figure 1, $A_{u^2}$ is a complete subgraph of $\Gamma(R)$ and hence the girth of $\Gamma(R)$ is 3.
\end{proof}

\begin{theorem}
The vertex connectivity $\kappa(\Gamma(R))$ of $\Gamma(R)$ is $p-1.$
\end{theorem}
\begin{proof}
The degree of any vertex in $\Gamma(R)$ is at least $p-1.$ Therefore, minimum $p-1$ vertices are removed from the graph to be disconnected. Hence, the vertex connectivity is $\kappa(\Gamma(R))=p-1.$
\end{proof}

\begin{theorem}
The edge connectivity $\lambda(\Gamma(R))$ of $\Gamma(R)$ is $p-1.$
\end{theorem}
\begin{proof}
As $\Gamma(R)$ connected graph, $\kappa(\Gamma(R))\leq \lambda(\Gamma(R))\leq \delta(\Gamma(R)).$ Since $\kappa(\Gamma(R))=p-1$ and $\delta(\Gamma(R))=p-1,$ then $\lambda(\Gamma(R))=p-1.$ 
\end{proof}

\section{Some Topological Indices of $\Gamma(R)$}
In this section, we find the Wiener index, first Zagreb index, second Zagreb index and Randi\'{c} index of the zero divisor graph $\Gamma(R).$

\begin{theorem}
The Wiener index of the zero-divisor graph $\Gamma(R)$ of $R$ is
$W(\Gamma(R))=\frac{p(2p^3-2p^2-7p+5)}{2}.$
\end{theorem}
\begin{proof}
Consider,
\begin{align*}
    W(\Gamma(R))&=\sum_{\substack{x, y\in Z^{*}(R)\\ x\neq y}} d(x, y)\\
    &=\sum\limits_{\substack{x, y\in A_{u}\\ x\neq y}} d(x, y)+\sum\limits_{\substack{x, y\in A_{u^2}\\ x\neq y}} d(x, y)+\sum\limits_{\substack{x, y\in A_{u+u^2}\\ x\neq y}} d(x, y)\\
    &\hspace{1cm}+\sum\limits_{\substack{x\in A_u\\ y\in A_{u^2}}} d(x, y)+\sum\limits_{\substack{x\in A_{u}\\ y\in A_{u+u^2}}} d(x, y)+\sum\limits_{\substack{x\in A_{u^2}\\ y\in A_{u+u^2}}} d(x, y)\\
    &=(p-1)(p-2)+\frac{(p-1)(p-2)}{2}+p(p-2)(p-1)^2\\
    &\hspace{1cm}+(p-1)^2+2(p-1)^3+ (p-1)^3\\
    &=(p-1)^2+3(p-1)^3+\frac{(p-1)(p-2)}{2}+(p-1)(p-2)(p^2-p+1)\\
    &=\frac{p(2p^3-2p^2-7p+5)}{2}.
\end{align*}
\end{proof}
Denote $[A, B]$ be the set of edges between the subset $A$ and $B$ of $V.$ For any $a\in A_u, d_a=p-1,$ for any $a\in A_{u^2}, d_a=p^2-2$ and any $a\in A_{u+u^2}, d_a=p-1.$
\begin{theorem}
The Randi\'{c} index of the zero-divisor graph $\Gamma(R)$ of $R$ is
$R(\Gamma(R))=\frac{(p-1)}{2(p^2-2)}\Big[2p\sqrt{(p-1)(p^2-2)}+(p-2)\Big].$
\end{theorem}
\begin{proof}
Consider,
\begin{align*}
    R(\Gamma(R))&=\sum\limits_{(a,b)\in E} \frac{1}{\sqrt{d_a d_b}}\\
    &=\sum\limits_{(a,b)\in [A_u, A_{u^2}]} \frac{1}{\sqrt{d_a d_b}}+\sum\limits_{(a,b)\in [A_{u^2}, A_{u^2}]} \frac{1}{\sqrt{d_a d_b}}+\sum\limits_{(a,b)\in [A_{u^2}, A_{u+u^2}]} \frac{1}{\sqrt{d_a d_b}}\\
    &=(p-1)^2 \frac{1}{\sqrt{(p-1)(p^2-2)}}
    +\frac{(p-1)(p-2)}{2} \frac{1}{\sqrt{(p^2-2)(p^2-2)}}\\
    &\hspace{1cm}+(p-1)^3\frac{1}{\sqrt{(p^2-2)(p-1)}}\\
    &=\frac{(p-1)^2}{\sqrt{(p-1)(p-2)}}[p(p-1)]+\frac{(p-1)(p-2)}{2(p^2-2)}\\
    &=\frac{p(p-1)^2}{\sqrt{(p-1)(p^2-2)}}+\frac{(p-1)(p-2)}{2(p^2-2)}\\
    &=\frac{(p-1)}{2(p^2-2)}\Big[2p\sqrt{(p-1)(p^2-2)}+(p-2)\Big]
    \end{align*}
\end{proof}

\begin{theorem}
The first Zagreb index of the zero-divisor graph $\Gamma(R)$ of $R$ is
$M_1(\Gamma(R))=(p-1)[p^4+p^3-4p^2+p+4].$
\end{theorem}
\begin{proof}
Consider,
\begin{align*}
    M_1(\Gamma(R))&=\sum\limits_{a\in Z^{*}(R)} d_a^2\\
    &=\sum_{a\in A_u}d_a^2+\sum_{a\in A_{u^2}}d_a^2+\sum_{a\in A_{u+u^2}}d_a^2\\
    &=(p-1)(p-1)^2+(p-1)(p^2-2)^2+(p-1)^2(p-1)^2\\
    &=(p-1)^3+(p-1)^4+(p^2-2)^2(p-1)\\
    &=p(p-1)^3+(p-1)(p^2-2)\\
    &=(p-1)[p^4+p^3-4p^2+p+4].
    \end{align*}
\end{proof}
\begin{theorem}
The second Zagreb index of the zero-divisor graph $\Gamma(R)$ of $R$ is
$M_2(\Gamma(R))=\frac{1}{2}[3p^6-9p^5+22p^3-16p^2-8p+8].$
\end{theorem}
\begin{proof}
Consider,
\begin{align*}
    M_2(\Gamma(R))&=\sum\limits_{(a,b)\in E} d_a d_b\\
    &=\sum\limits_{(a,b)\in [A_u, A_{u^2}]} d_a d_b+\sum\limits_{(a,b)\in [A_{u^2}, A_{u^2}]} d_a d_b+\sum\limits_{(a,b)\in [A_{u^2}, A_{u+u^2}]} d_a d_b\\
    &=(p-1)^2(p-1)(p^2-2)
    +\frac{(p-1)(p-2)}{2} (p^2-2)(p^2-2)\\
    &\hspace{2cm}+(p-1)^3(p^2-2)(p-1)\\
    &=\dfrac{(p-1)(p^2-2)}{2}[3p^3-6p^2+4]\\
    &=\frac{1}{2}[3p^6-9p^5+22p^3-16p^2-8p+8].
    \end{align*}
\end{proof}

\section{Codes from Incidence Matrix of $\Gamma(R)$}
In this section, we find the incidence matrix of the graph $\Gamma(R)$ and we find the parameters of the linear code generated by the rows of incidence matrix $Q(\Gamma(R)).$ 

 The incidence matrix $Q(\Gamma(R))$ is given below
$$Q(\Gamma(R)) = \bordermatrix{~ & [A_u, A_{u^2}] &[A_{u^2}, A_{u^2}] & [A_{u^2},A_{u+u^2}]\cr
                  A_{u} & D^{(p-1)}_{(p-1)\times (p-1)^2} & {\bf 0}_{(p-1)\times \frac{(p-1)(p-2)}{2}} &{\bf 0}_{(p-1)\times (p-1)^3}\cr
                   A_{u^2} & J_{(p-1)\times (p-1)^2} & J_{(p-1)\times \frac{(p-1)(p-2)}{2}}&J_{(p-1)\times (p-1)^3} \cr
                  A_{u+u^2} & {\bf 0}_{(p-1)^2\times (p-1)^2} & {\bf 0}_{(p-1)^2\times\frac{(p-1)(p-2)}{2}}& D^{(p-1)}_{(p-1)^2\times (p-1)^3}},$$
where $J$ is a all one matrix, ${\bf 0}$ is a zero matrix with appropriate order, ${\bf 1}_{(p-1)}$ is a all one $1\times (p-1)$ row vector 
and $D^{(p-1)}_{k\times l}=\begin{pmatrix}
{\bf 1}_{(p-1)}&{\bf 0}&{\bf 0}&\dots&{\bf 0}\\
{\bf 0}&{\bf 1}_{(p-1)}&{\bf 0}&\dots&{\bf 0}\\
\vdots&\vdots&\vdots&\ddots&\vdots\\
{\bf 0}&{\bf 0}&{\bf 0}&\dots&{\bf 1}_{(p-1)}
\end{pmatrix}_{k\times l}.$

\begin{example}\label{b}
The incidence matrix of the zero-divisor graph $\Gamma(R)$ given in the Example \ref{a} is
$$Q(\Gamma(R))=\begin{matrix}u\\2u\\ u^2\\2u^2\\ u+u^2\\2u+2u^2\\2u+u^2\\u+2u^2\end{matrix}\\\left(\begin{array}{rrrrrrrrrrrrrrr}
1 & 1 & 0&0&\vline&0&\vline&0& 0&0&0&0&0&0&0\\
0 & 0 & 1&1&\vline&0&\vline&0& 0&0&0&0&0&0&0\\
\hline
  1 & 1 & 1&1&\vline&1&\vline&1&1&1&1&1&1&1&1\\
  1 & 1 & 1&1&\vline&1&\vline&1&1&1&1&1&1&1&1\\
  \hline
    0 & 0 & 0&0&\vline&0&\vline&1& 1&0&0&0&0&0&0\\
   0 & 0 & 0&0&\vline&0&\vline&0& 0&1&1&0&0&0&0 \\
   0 & 0 &  0&0&\vline&0&\vline&0& 0&0&0&1&1&0&0\\
  0 & 0 &  0&0&\vline&0&\vline&0& 0&0&0&0&0& 1&1
\end{array}\right)_{8\times 13}.$$
The number of linearly independent rows is 7 and hence the rank of the matrix $Q(\Gamma(R))$ is 7.
The rows of the incidence matrix $Q(\Gamma(R))$ is  generate a $[n=13, k=7, d=2]_2$ code over $\mathbb{F}_2.$ 
\end{example}

The edge connectivity of the zero-divisor graph $\Gamma(R)$ is $p-1,$ then we have the following theorem:
\begin{theorem}
The linear code generated by the incidence matrix $Q(\Gamma(R))$ of the zero-divisor graph $\Gamma(R)$ is a  $C_2(\Gamma(R))=[\frac{1}{2}(2p^3-3p^2-p+2), p^2-2, p-1]_2$  linear code over the finite field $\mathbb{F}_2.$
\end{theorem}
 
\section{Adjacency and Laplacian Matrices of $\Gamma(R)$}
In this section, we find the eigenvalues, energy and spectral radius of both adjacency and Laplacian matrices of $\Gamma(R)$.

If $\mu$ is an eigenvalue of matrix $A$ then  $\mu^{(k)}$ means that $\mu$ is an eigenvalue with multiplicity $k.$ 

The vertex set partition into $A_u, A_{u^2}$ and $A_{u+u^2}$ of cardinality $p-1,p-1$ and $(p-1)^2,$ respectively.
Then the adjacency matrix of $\Gamma(R)$ is

$$A(\Gamma(R)) = \bordermatrix{~ & A_u & A_{u^2} & A_{u+u^2}\cr
    A_u&{\bf0}_{p-1} & J_{p-1} & {\bf0}_{(p-1)\times (p-1)^2}\cr
  A_{u^2}&J_{p-1} & J_{p-1}-I_{p-1} & J_{(p-1)\times (p-1)^2}\cr
  A_{u+u^2}&{\bf 0}_{(p-1)^2\times (p-1)} & J_{(p-1)^2\times (p-1)} & {\bf 0}_{(p-1)^2} },$$

where $J_k$ is an $k\times k$ all one matrix, $J_{n\times m}$ is an $n\times m$ all matrix, ${\bf 0}_{k}$ is an $k\times k$ zero matrix, ${\bf 0}_{n\times m}$ is an $n\times m$ zero matrix and $I_{k}$ is an $k\times k$ identity matrix.

All the rows in $A_{u^2}$ are linearly independent and all the rows in $A_u$ and $A_{u+u^2}$ are linearly dependent. Therefore, $p-1+1=p$ rows are linearly independent. So, the rank of $A(\Gamma(R))$ is $p.$ By Rank-Nullity theorem, nullity of $A(\Gamma(R))=p^2-p-1.$ Hence, zero is an eigenvalue with multiplicity $p^2-p-1.$

For $p=3,$ the adjacency matrix of $\Gamma(R)$ is
$$A(\Gamma(R))=\left(\begin{array}{rrrrrrrrrr}
0 & 0 &\vline& 1 & 1 &\vline& 0 & 0 & 0 & 0 \\
0 & 0 &\vline& 1 & 1 &\vline& 0 & 0 & 0 & 0 \\
\hline
1 & 1 &\vline& 0 & 1 &\vline& 1 & 1 & 1 & 1 \\
1 & 1 &\vline& 1 & 0 &\vline& 1 & 1 & 1 & 1 \\
\hline
0 & 0 &\vline& 1 & 1 &\vline& 0 & 0 & 0 & 0 \\
0 & 0 &\vline& 1 & 1 &\vline& 0 & 0 & 0 & 0 \\
0 & 0 &\vline& 1 & 1 &\vline& 0 & 0 & 0 & 0 \\
0 & 0 &\vline& 1 & 1 &\vline& 0 & 0 & 0 & 0
\end{array}\right)_{8\times 8}.$$
The eigenvalues of $A(\Gamma(R))$ are $0^{(5)}, 4^{(1)}, (-1)^{(1)}$ and $ (-3)^{(1)}.$
For $p=5,$ the eigenvalues  of $A(\Gamma(R))$ are $0^{(19)}, 10^{(1)}, (-1)^{(3)}$ and $ (-7)^{(1)}.$  
\begin{theorem}
The energy of the adjacency matrix $A(\Gamma(R))$ is $\varepsilon(\Gamma(R))=6p-10.$
\end{theorem}
\begin{proof}
For any odd prime $p,$ the eigenvalues of $A(\Gamma(R))$ are $0^{(p^2-p-1)},$ $(3p-5)^{(1)},$ $(-1)^{(p-2)},$ $(3-2p)^{(1)}.$
The energy of adjacency matrix $A(\Gamma(R))$ is the sum of the absolute values of all eigenvalues of $A(\Gamma(R)).$
That is,
\begin{align*}
    \varepsilon(\Gamma(R))&=\sum_{i=1}^{p^2-1}|\lambda_i|~~~~~~ \text{ where $\lambda_i$'s are eigenvalues of $A(\Gamma(R))$} \\
    &=|3p-5|+(p-2)|-1|+|3-2p|\\
    &=3p-5+p-2+2p-3~~~~~~\text{ since } p>2\\
    &=6p-10.
\end{align*}
\end{proof}
The degree matrix of the graph $\Gamma(R)$ is
$$D(\Gamma(R)) = \bordermatrix{~ & A_u & A_{u^2} & A_{u+u^2}\cr
    A_u&(p-1)I_{p-1} & {\bf 0}_{p-1} & {\bf0}_{(p-1)\times (p-1)^2}\cr
  A_{u^2}&{\bf 0}_{p-1} & (p^2-2)I_{p-1} & {\bf 0}_{(p-1)\times (p-1)^2}\cr
  A_{u+u^2}&{\bf 0}_{(p-1)^2\times (p-1)} & {\bf 0}_{(p-1)^2\times (p-1)} & (p-1)I_{(p-1)^2} }.$$
The Laplacian matrix $L(\Gamma(R))$ of $\Gamma(R)$ is defined by $L(\Gamma(R))=D(\Gamma(R))-A(\Gamma(R)).$ Therefore,

$$L(\Gamma(R)) = \bordermatrix{~ & A_u & A_{u^2} & A_{u+u^2}\cr
    A_u&(p-1)I_{p-1} & -J_{p-1} & {\bf0}_{(p-1)\times (p-1)^2}\cr
  A_{u^2}&-J_{p-1} & (p^2-1)I_{p-1}-J_{p-1} & -J_{(p-1)\times (p-1)^2}\cr
  A_{u+u^2}&{\bf 0}_{(p-1)^2\times (p-1)} & -J_{(p-1)^2\times (p-1)} & (p-1)I_{(p-1)^2} }.$$
Since each row sum is zero, zero is one of the eigenvalues of $L(\Gamma(R)).$ By Lemma \ref{a}, the second smallest eigenvalue of $L(\Gamma(R))$ is positive as $\Gamma(R)$ is connected. Hence zero is an eigenvalue with multiplicity one, and all other eigenvalues are positive.

For $p=3,$ the Laplacian matrix is
$$L(\Gamma(R))=\left(\begin{array}{rrrrrrrr}
2 & 0 & -1 & -1 & 0 & 0 & 0 & 0 \\
0 & 2 & -1 & -1 & 0 & 0 & 0 & 0 \\
-1 & -1 & 7 & -1 & -1 & -1 & -1 & -1 \\
-1 & -1 & -1 & 7 & -1 & -1 & -1 & -1 \\
0 & 0 & -1 & -1 & 2 & 0 & 0 & 0 \\
0 & 0 & -1 & -1 & 0 & 2 & 0 & 0 \\
0 & 0 & -1 & -1 & 0 & 0 & 2 & 0 \\
0 & 0 & -1 & -1 & 0 & 0 & 0 & 2
\end{array}\right)_{8\times 8}.$$
The eigenvalues of $L(\Gamma(R))$ are $0^{(1)}, 8^{(2)}, 2^{(5)}.$

For $p=5,$ the eigenvalues of $L(\Gamma(R))$ are  $0^{(1)}, 24^{(4)}, 4^{(19)}.$

For any prime $p,$ the eigenvalues of $L(\Gamma(R))$ are $0^{(1)}, (p^2-1)^{(p-1)}, (p-1)^{(p^2-p-1)}.$

\begin{theorem}
The Laplacian energy of $\Gamma(R)$ is $LE(\Gamma(R))=\dfrac{2p^5-6p^4+6p^3-4p+1}{p^2-1}.$
\end{theorem}
\begin{proof}
Let $|V|=n$ and $|E|=m.$  Let $\mu_1, \mu_2,\dots,\mu_n$ are eigenvalues of $L(\Gamma(R)).$ Then
the Laplacian energy $LE(\Gamma(R))$ is given by $$LE(\Gamma(R))=\sum\limits_{i=1}^n \Big|\mu_i-\frac{2m}{n}\Big|.$$
We know that the eigenvalues of $L(\Gamma(R))$ are $0^{(1)}, (p^2-1)^{(p-1)}, (p-1)^{(p^2-p-1)}.$
Then
\begin{align*}
    LE(\Gamma(R))&=\sum\limits_{i=1}^n \Big|\mu_i-\frac{2m}{n}\Big|\\
    &=\sum\limits_{i=1}^n \Big|\mu_i-\frac{2p^3-3p^2-p+2}{p^2-1}\Big|\\
    &=\Big|0-\frac{2p^3-3p^2-p+2}{p^2-1}\Big|+(p-1)\Big|(p^2-1)-\frac{2p^3-3p^2-p+2}{p^2-1}\Big|\\
    &\hspace{2.5cm}+(p^2-p-1)\Big|(p-1)-\frac{2p^3-3p^2-p+2}{p^2-1}\Big|\\
    &=\frac{(2p^3-3p^2-p+2)+(p-1)\Big|p^4-2p^3+p^2+p-1\Big|+(p^2-p-1)\Big|p^3-2p^2+1\Big|}{p^2-1}\\
    &=\dfrac{2p^5-6p^4+6p^3-4p+1}{p^2-1}\hspace{1cm} \text{ since } p\geq 2.
\end{align*}
\end{proof}
We denote by $\rho(\Gamma(R))$ the largest
eigenvalue in absolute of $A(\Gamma(R))$ and call it the spectral radius of $\Gamma(R);$ we denote by $\mu(\Gamma(R))$ the largest eigenvalue in absolute of $L(\Gamma(R))$ and call it the Laplacian spectral radius of $\Gamma(R).$
\begin{theorem}
 For any odd prime $p,$  $\rho(\Gamma(R))=3p-5$ and  $\mu(\Gamma(R))=p^2-1.$
\end{theorem}
\begin{proof}
The eigenvalues of the adjacency matrix $A(\Gamma(R))$ are $0^{(p^2-p-1)},$ $(3p-5)^{(1)},$ $(-1)^{(p-2)}$ and $(3-2p)^{(1)}.$ Then the largest eigenvalue in absolute is $3p-5$ as $p>2.$ That is, $\rho(\Gamma(R))=3p-5.$

The eigenvalues of the Laplacian matrix $L(\Gamma(R))$ are $0^{(1)}, \,\, (p^2-1)^{(p-1)}$ and $(p-1)^{(p^2-p-1)}.$  Then the largest eigenvalue in absolute is $p^2-1.$  That is, $\mu(\Gamma(R))=p^2-1.$

\end{proof}
\section*{Conclusion}
In this article, we discussed the zero-divisor graph of a commutative ring with identity $\mathbb{F}_p+u\mathbb{F}_p+u^2 \mathbb{F}_p$ where $u^3=0$ and $p$ is an odd prime.
We find the clique number, chromatic number, vertex connectivity, edge connectivity, diameter and girth of a zero-divisor graph associated with the ring. 
We find some of topological indices and the main parameters of the code derived from the incidence matrix of the zero-divisor graph $\Gamma(R).$ 
Also, we find the eigenvalues, energy and spectral radius  of both adjacency and Laplacian matrices of $\Gamma(R).$ 

\section*{Acknowledgements}
The author thanks Satya Bagchi and C. Durairajan  for their help to improving the presentation of the paper.

\end{document}